\documentclass[aps,prl,reprint,amsmath,amssymb,superscriptaddress, nofootinbib]{revtex4-1}
\usepackage[utf8]{inputenc}
\usepackage{hepunits}
\usepackage{graphicx}
\usepackage{dcolumn}
\usepackage{bm}
\usepackage[mathlines]{lineno}
\usepackage{lipsum}
\usepackage{xspace}
\usepackage[nolist]{acronym}
\usepackage{color}
\usepackage{subfigure}

\graphicspath{{images/}}

\begin{document}
\newcommand{\kgyr}{kg$\cdot$yr\xspace}
\newcommand{\cky}{counts/(kg$\cdot$yr)\xspace}
\newcommand{\ckky}{counts/(keV$\cdot$kg$\cdot$yr)\xspace}
\newcommand{\TeO}{TeO$_2$\xspace}
\newcommand{\isoTe}{$^{130}$Te\xspace}
\newcommand{\NDBD}{$0\nu\beta\beta$ decay\xspace}
\newcommand{\DBD}{$\nu\beta\beta$ decay\xspace}
\newcommand{\nbb}{$\nu\beta\beta$\xspace}
\newcommand{\Qbb}{$Q_{\beta\beta}$\xspace}
\newcommand{\qino}{Cuoricino\xspace}
\newcommand{\q}{\mbox{CUORE-0}\xspace}
\newcommand{\GEANT}{G{\sc eant}}

\title[Recent Progress of the CUORE Experiment]{Update on the recent progress of the CUORE experiment\\
(Proceedings of the Neutrino 2018 Conference)}
\author{D.~Q.~Adams}
\affiliation{Department of Physics and Astronomy, University of South Carolina, Columbia, SC 29208, USA}

\author{C.~Alduino}
\affiliation{Department of Physics and Astronomy, University of South Carolina, Columbia, SC 29208, USA}

\author{K.~Alfonso}
\affiliation{Department of Physics and Astronomy, University of California, Los Angeles, CA 90095, USA}

\author{F.~T.~Avignone~III}
\affiliation{Department of Physics and Astronomy, University of South Carolina, Columbia, SC 29208, USA}

\author{O.~Azzolini}
\affiliation{INFN -- Laboratori Nazionali di Legnaro, Legnaro (Padova) I-35020, Italy}

\author{G.~Bari}
\affiliation{INFN -- Sezione di Bologna, Bologna I-40127, Italy}

\author{F.~Bellini}
\affiliation{Dipartimento di Fisica, Sapienza Universit\`{a} di Roma, Roma I-00185, Italy}
\affiliation{INFN -- Sezione di Roma, Roma I-00185, Italy}

\author{G.~Benato}
\affiliation{Department of Physics, University of California, Berkeley, CA 94720, USA}

\author{A.~Bersani}
\affiliation{INFN -- Sezione di Genova, Genova I-16146, Italy}

\author{M.~Biassoni}
\affiliation{INFN -- Sezione di Milano Bicocca, Milano I-20126, Italy}

% ...
\date{August 27, 2018}

\begin{abstract}
  CUORE is a 741\,kg array of 988 \TeO bolometeric crystals designed
  to search for the neutrinoless double beta decay of $^{130}$Te and
  other rare processes. CUORE has been taking data since summer 2017,
  and as of summer 2018 collected a total of 86.3\,\kgyr of \TeO
  exposure. Based on this exposure, we were able to set a limit on the
  \NDBD half-life of \isoTe of $T^{0\nu}_{1/2}>1.5\times10^{25}$\,yr
  at 90\% C.L. At this conference, we showed the decomposition of the
  CUORE background and were able to extract a \isoTe 2\nbb half-life
  of
  $T_{1/2}^{2\nu}=[7.9\pm0.1\,\mathrm{(stat.)}\pm0.2\,\mathrm{(syst.)}]\times10^{20}$\,yr. This
  is the most precise measurement of this half-life and is consistent
  with previous measurements.
\end{abstract}

\maketitle
\begin{acronym}
\acro{CUORE}{Cryogenic Underground Observatory for Rare Events}
\acro{NDBD}[0$\nu\beta\beta$ decay]{Neutrinoless Double Beta decay}
\acro{SM}{Standard Model}
\acro{NTD}{neutron transmutation doped thermistor}
\acro{FWHM}{full-width half-max}
\acro{MC}{Monte Carlo}
\acro{MCMC}{Markov-Chain Monte Carlo}
\acro{ROI}{region of interest}
\end{acronym}

\newcommand{\Q}{\ac{CUORE}\xspace}
\renewcommand{\NDBD}{\ac{NDBD}\xspace}
\newcommand{\SM}{\ac{SM}\xspace}
\newcommand{\NTD}{\ac{NTD}\xspace}
\newcommand{\FWHM}{\ac{FWHM}\xspace}
\newcommand{\MC}{\ac{MC}\xspace}
\newcommand{\MCMC}{\ac{MCMC}\xspace}
\newcommand{\ROI}{\ac{ROI}\xspace}

\section{Introduction}

In the early 2000s, the discovery of neutrino oscillations
demonstrated conclusively that neutrinos are massive particles
\cite{Fukuda2002,Ahmad2001,Ahmad2002,Eguchi2003,Gando2011,PDG2018}. But
despite the nearly 20 years that have passed since this discovery, the
scale and nature of that mass has still not been established. Whether
the neutrino has a Dirac type mass or a Majorana type mass is one of
the most actively sought questions in particle physics today. A
Majorana type mass would indicate the violation of lepton number,
create a natural explanation for the neutrino's lightness, point
towards new physics well beyond the scale of today's accelerators, and
would even have implications for the formation of the matter asymmetry
of the universe.

The \Q is primarily a search for the \NDBD of
\mbox{$^{130}\mathrm{Te}\rightarrow^{130}\mathrm{Xe}+2e^{-}$}
\cite{CUORE_arxiv}. The discovery of this decay would indicate
conclusively that the neutrino is a Majorana fermion. \Q builds on a
long history of cryogenic searches for the \NDBD of \isoTe
\cite{Q0FinalPrl,Q0FinalPRC,QINOPaper,Cuoricino_PLB,Cuoricino_PRC} and
is the first such experiment to reach the ton scale. The search for
\NDBD is a very active area of research, with multiple experiments
searching in a variety of candidate isotopes using a range of detector
technologies \cite{GERDA2018,KamLANDZen2016,EXO2018}.

\begin{figure*}
  \centering                                                                 
  \hfill
  \begin{minipage}{.155\textwidth}
    \subfigure[]{\label{fig:tower}\includegraphics[width=0.45\linewidth]{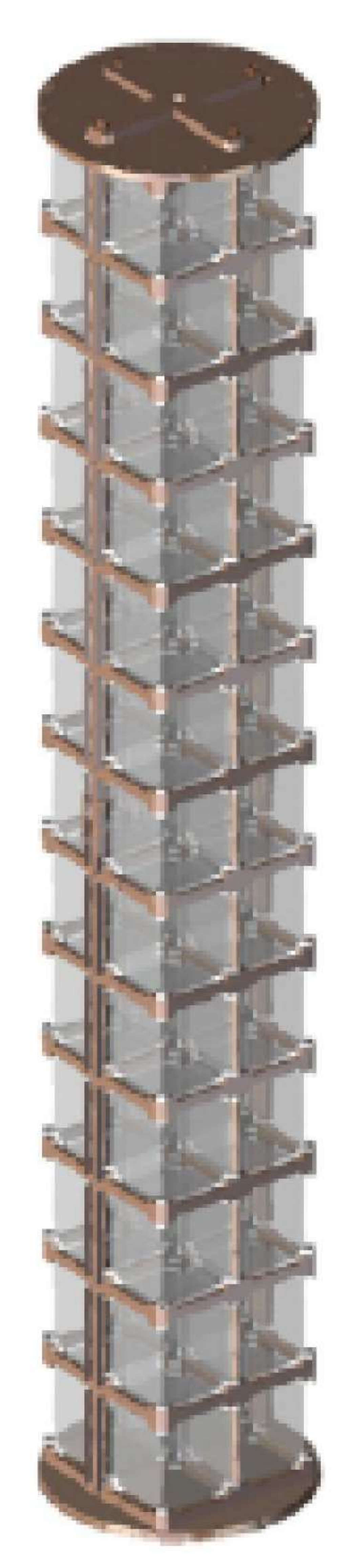}}
  \end{minipage}
  \begin{minipage}{.83\textwidth}
    \subfigure[]{\label{fig:bolo_cartoon}%
      \includegraphics[width=0.4\linewidth]{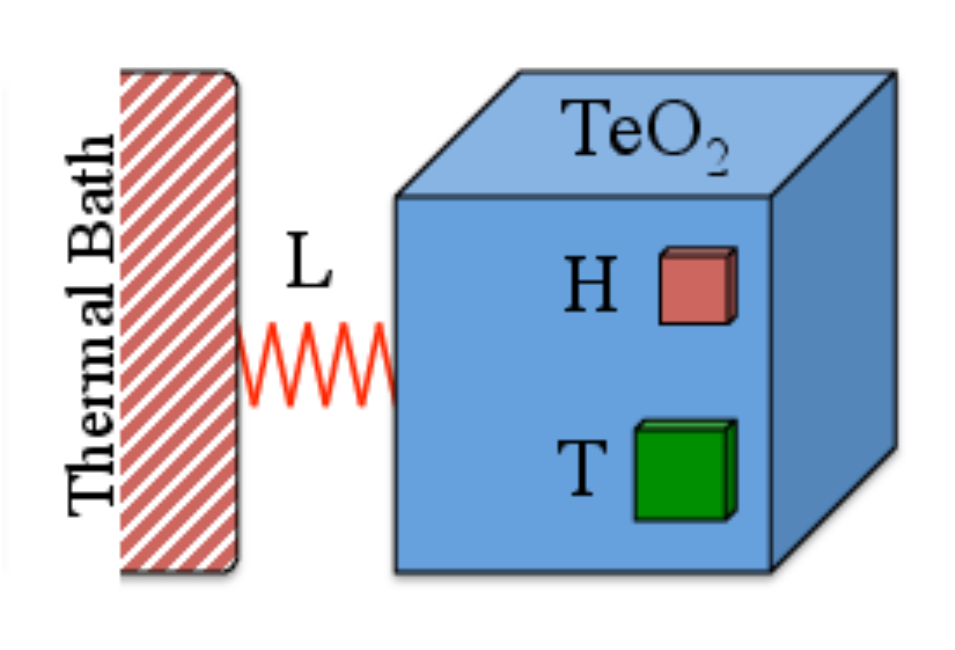}%
    }
    \hspace{1cm}
    \subfigure[]{\label{fig:installed_detector}%
      \includegraphics[width=0.45\linewidth]{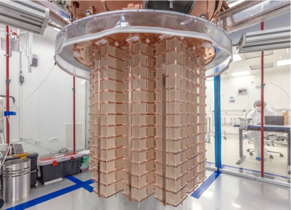}%
    }
  \end{minipage}
  \hfill
  \caption{(a) Schematic of a single \Q-like tower, with 13 floors of
    4 crystals each. Each tower has 52 crystals for a total mass of
    $\sim$39\,kg of \TeO or $\sim$10.8\,kg of \isoTe. The \Q detector
    is composed of 19 such towers. (b) A schematic of a CUORE
    bolometer. The \TeO crystal acts as the absorber, and is connected
    to a heat bath through a weak thermal link, \emph{L}. Each
    bolometer is instrumented with an NTD thermistor, \emph{N}, and a
    heater to inject heat pulses, \emph{H}. (c) The \Q detector
    assembled and installed in the \Q cryostat, inside the \Q
    cleanroom.}
\end{figure*}

The \Q detector is an array of 988 bolometers operating independently,
with each bolometer acting as an individual search for \NDBD. A CUORE
bolometer is composed of two main components, an absorber which
absorbs the energy released in a particle interaction and converts
that energy into an increase in temperature and a thermistor which
converts this change in temperature into a measurable change in
voltage (see Fig.~\ref{fig:bolo_cartoon}). A CUORE absorber is a \TeO
crystal, 5$\times$5$\times$5 cm$^3$ in size and weighing approximately
750~g. The crystal is made using $^{\rm nat}$Te, which is $\sim$34\%
$^{130}$Te, and thus acts as both the source and detector of the
decays of interest. When the decay occurs, the emitted electrons
deposit their energy into the crystal lattice in the form of heat and
any energy imparted to neutrinos is lost to the detector. Since no
neutrinos are produced in \NDBD, the signal that we are searching for
is a narrow peak at the full energy of the \isoTe decay,
$Q_{\beta\beta}=2527.5$\,keV.

The \Q crystals are arranged into an array of 19 towers, 13 floors
each, with 4 crystals per floor (Fig.~\ref{fig:tower}). Each crystal
has a mass of $\sim$750\,g, giving the \Q detector a total active mass
of 741\,kg, of which $\sim$206\,kg is $^{130}$Te. This
\mbox{source\,=\,detector} configuration gives us a high signal
efficiency of around $\gtrsim85$\%. The detector is cooled to its
operating temperature of $\sim$10\,mK in the \Q cryostat
(Fig.~\ref{fig:installed_detector}). At this low temperature, the heat
capacity of a crystal falls to the point that a 1\,MeV energy
deposition causes a change in temperature of about 100~$\mu$K. This
small change in temperature is read out using an \NTD with a
resistivity that is exponentially dependent on temperature and
amplifies the $\sim$1\% temperature change to a change in resistivity
of $\sim$10\%. The NTDs have typical resistances of
$\sim0.1-1$\,G$\Omega$ and are current biased and read out using room
temperature electronics.

\section{Background Reduction}
The figure of merit that determines the \Q sensitivity to \NDBD can be
expressed as \mbox{$\propto\varepsilon \sqrt{M T/(b\Delta E)}$}
\cite{SensitivityPaper}, where $\varepsilon$ is the total signal
efficiency, $M$ is the active mass, $T$ the total detector live time,
and $\Delta E$ and $b$ are the energy resolution in keV and background
in \ckky at \Qbb. \Q typically aims for an energy resolution of about
$\Delta E \approx 5$\, keV \FWHM at \Qbb.

The sensitivity to \NDBD is thus driven by the background index, $b$,
and a lot of work has been done over the years to improve it. The
primary backgrounds in \Q come from naturally occurring radioactivity
that originates on or near materials close to our bolometers. The
predecessor to \Q, called \qino
\cite{QINOPaper,Cuoricino_PLB,Cuoricino_PRC}, had a background index
of \mbox{$b=0.169\pm0.006$\,\ckky}, of which,
\mbox{$0.110\pm0.001$\,\ckky} came from $\alpha$ particles originating
on surfaces of the bolometers or materials immediately facing the
bolometers. The remaining background came from $\gamma$-rays
originating in materials around the detector (but necessarily
immediately facing it). \qino ran from 2003 -- 2008 and with
19.75\,\kgyr of \isoTe exposure was able to set a limit on the \NDBD
half-life of \isoTe of \mbox{$T_{1/2}^{0\nu} > 2.8 \times 10^{24}$\,yr} at
90\%~C.L.

The first phase of \Q was a single tower experiment called \q
\cite{Q0Detector,Q0InitialPaper} operated in the same cryostat that
housed \qino. \q was designed to be a full scale test of the new
surface cleaning and detector assembly procedures developed to reduce
the $\alpha$-background observed in \qino
\cite{Alessandria2013,CUOREAssemblyLine}. \q was able to achieve a
background index at \Qbb of
$b=0.058\pm0.004\mathrm{(stat.)}\pm0.002\mathrm{(syst.)}$\,\ckky with
only $0.016\pm0.001$\,\ckky coming from $\alpha$-backgrounds. The
$\gamma$-background was observed to be consistent with \qino as was
anticipated since the $\gamma$ backgrounds were believed to originate
in the materials of the cryostat, which was common to both
experiments. \q ran from 2013 -- 2015 and collected an exposure of
9.8\,\kgyr of \isoTe. Using the result of \qino as a prior, \q
set a limit on the \NDBD of \isoTe of
$T_{1/2}^{0\nu}>4.0\times10^{24}$\,yr at
90\%~C.L. \cite{Q0FinalPrl,Q0FinalPRC}.

\Q uses the same surface cleaning and assembly procedure as \q and
aims to achieve a background at \Qbb of $b=0.01$\,\ckky
\cite{CUOREBudget}. The reduction in $\alpha$-background was mostly
demonstrated in \q, with the remaining reduction coming from the
larger volume-to-surface ratio of \Q. The $\gamma$-background is
reduced to a subdominant level by the cleaner materials used to build
the \Q cryostat, as well as the improved shielded and better
anti-coincidence capabilities of a larger detector.

\section{CUORE Data Taking}

\begin{figure}
  \centering                                                                   
  \hfill
  \begin{minipage}{.49\textwidth}
    \subfigure[]{\label{fig:calibrationFit}\includegraphics[width=\linewidth]{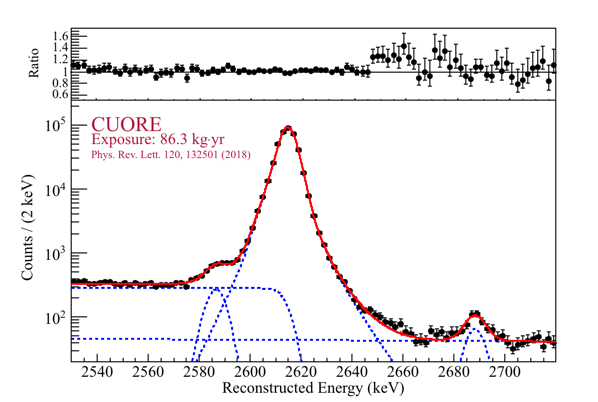}}
  \end{minipage}
  \hfill  
  \begin{minipage}{.49\textwidth}
    \subfigure[]{\label{fig:ROIFit}\includegraphics[width=\linewidth]{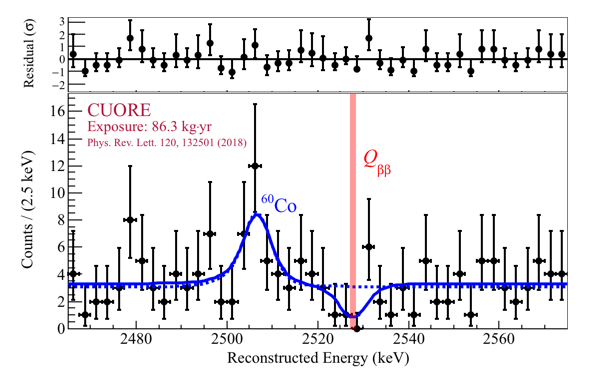}}
  \end{minipage}
  \hfill  
  \caption{(a) Reconstructed $^{208}$Tl 2516\,keV in the calibration
    spectrum. \emph{Lower}: The black points are data and the red line
    is the reconstructed line shape. The dotted blue lines are the
    components of the line shape corresponding to different physical
    processes (.e.g) full energy deposition, low angle
    compton-scattering, X-ray escape, etc. (b) The best fit in the
    ROI, with the expected \Qbb value at 2527.518\,keV marked. Both
    plots taken from \cite{CUOREPRL2017}.}
\end{figure}

The \Q detector completed assembly in summer 2014 and commissioning of
the \Q cryostat completed two years later in summer of 2016. The
detector was then installed into the cryostat inside a specially built
anti-radon tent; this minimized the exposure to radon during the
installation process\cite{CUOREAntiRadon}. The first \Q cooldown
started in December 2016 and reached base temperature in January 2017.

\begin{figure*}
\centering
  \begin{minipage}{\textwidth}
    \includegraphics[width=.95\textwidth]{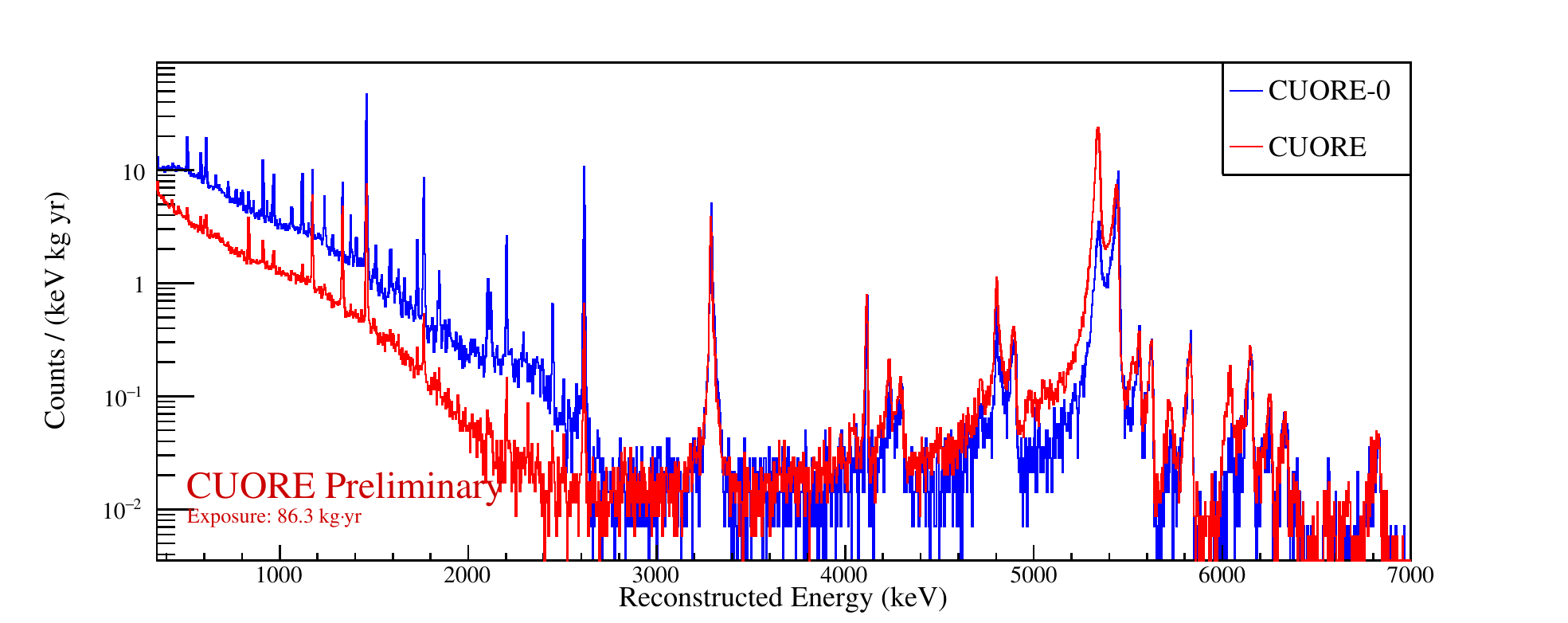}
    \caption{The observed CUORE M1 spectrum (\emph{red}) vs the
      CUORE-0 M1 spectrum (\emph{blue}). We see a significant decrease
      in the observed rate in the $\gamma$-region (below 3\,MeV). The
      $\alpha$-region (above 3\,MeV) is consistent with what was
      observed in \q. The excess of $^{210}$Po surface events can be
      see as the larger peak near 5.3\,MeV in the \Q
      spectrum. However, this does not appear to increase the
      background in the \NDBD ROI, as can be seen by the agreement of
      the backgrounds around 3\,MeV.}
  \label{fig:CUORE_vs_CUORE0}
  \end{minipage}
  \begin{minipage}{\textwidth}
    \includegraphics[width=.95\textwidth]{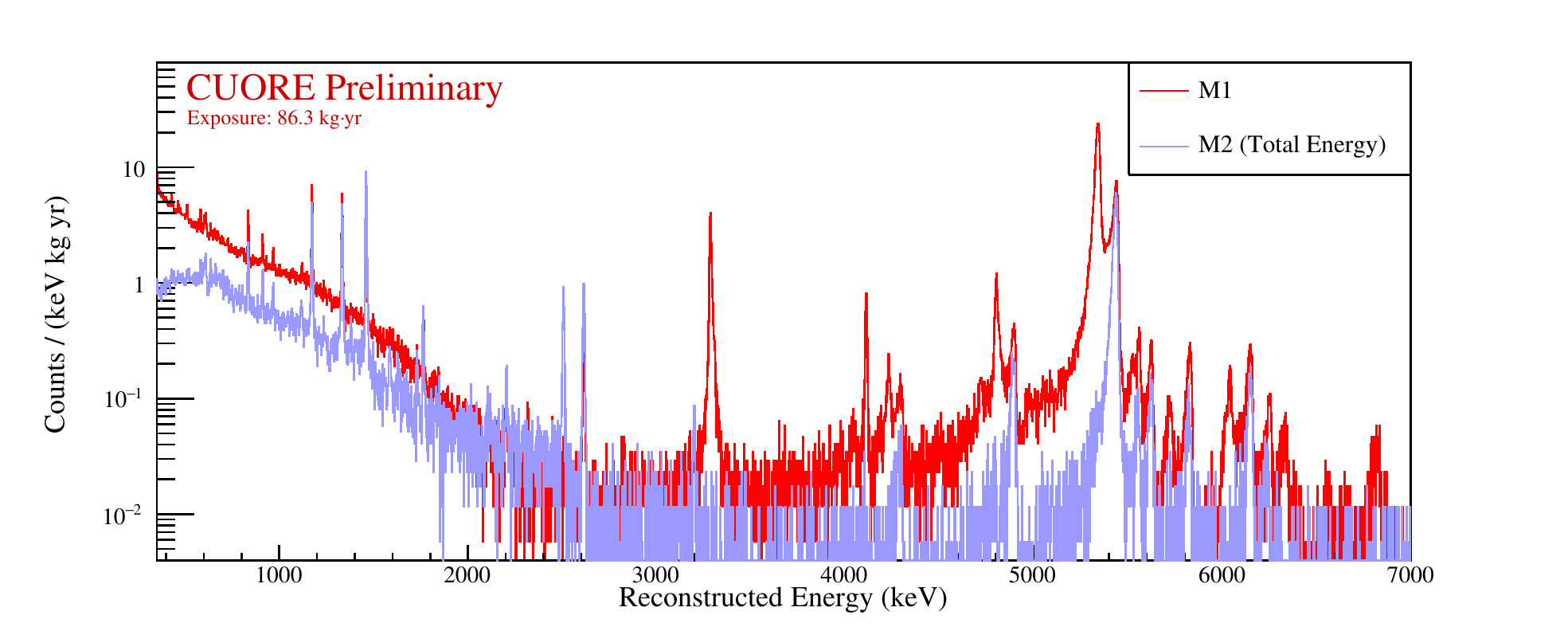}
    \caption{The \Q M1 spectrum vs the $\Sigma2$ spectrum. The
      $\Sigma2$ spectrum is more sensitive to $\gamma$ events and
      surface $\alpha$ events. We can see this by the lack of double
      peaks in the $\alpha$-region, since for an M2 event, the
      $\alpha$ and recoiling nucleus must both be detected. The peak
      at $\sim$3.2\,MeV corresponds to the $\alpha$-decay of
      $^{190}$Pt, which is a bulk contamination of the \TeO crystal
      from its growing procedure. Since the $^{190}$Pt is in the
      crystal bulk, the full energy is typically contained in the
      originating crystal and the peak only appears in the M1
      spectrum.}
    \label{fig:M1_vs_M2Sum}
  \end{minipage}
\end{figure*}

After a period of detector commissioning, \Q collected its first two
datasets in the summer of 2017. This data comprised a total of
86.3\,\kgyr of \TeO exposure (24.0\,\kgyr of \isoTe exposure)
characterized by an effective energy resolution of
$\Delta E = 7.7\pm0.5$\,keV \FWHM (see Fig.~\ref{fig:calibrationFit})
and a background index at \Qbb of $b=0.014\pm0.002$\,\ckky. The
background index is slightly higher than our background goal of
$0.01$\,\ckky in part because we have not yet implemented all of the
data quality cuts at our disposal. The energy resolution is also
larger than our goal of 5\,keV \FWHM. However, between the two
datasets presented here, we improved the energy resolution from
8.3\,keV to 7.4\,keV and we aim to continue to improve the resolution
as we continue to develop better algorithms as well as a better
understanding of the new \Q cryostat.

The summed spectrum from the first two datasets can be seen in
Fig.~\ref{fig:CUORE_vs_CUORE0} compared with the spectrum observed in
\q. Overall, the observed \Q spectrum was consistent with our
expectations. We saw a significant decrease in the $\gamma$-background
($\lesssim$3\,MeV) relative to \q. This is due to the significantly
improved material selection and shielding of the \Q cryostat. The
$\alpha$-region ($\gtrsim$3\,MeV) is consistent with what we observed
in \q. This was expected as \q and \Q shared the same surface cleaning
and assembly procedures. We do observe one unexpected background which
is an excess of surface $^{210}$Po events near 5.4\,MeV. The source of
this background is still unexplained, but the contamination appears to
be a very shallow surface contamination very close to the
bolometers. This contamination does not appear to significantly
increase the background in the \NDBD ROI (see
Fig.~\ref{fig:CUORE_vs_CUORE0}).

With the first two datasets, \Q was able to set a limit on the \NDBD
half-life of \isoTe of $T_{1/2}^{0\nu} > 1.3\times10^{25}$\,yr. This
result outperforms the expected half-life sensitivity of
$7.6\times10^{24}$\,yr due to a $\sim2\sigma$-downward fluctuation in
the background right at \Qbb. When combined with the results of \q and
\qino, we set the most stringent limit on the \NDBD half-life of
\isoTe to date at $T_{1/2}^{0\nu}>1.5\times10^{25}$\,yr at 90\%~C.L
\cite{CUOREPRL2017}. The ultimate sensitivity of CUORE after 5 years
of live time is $9\times10^{25}$\,yr.

\section{Measurement of the 2$\nu\beta\beta$ Decay Half-Life}

In order to understand the observed \Q spectrum, we simulate many
possible background sources with a \GEANT4 \cite{Geant} \MC
simulation. These backgrounds include surface contaminations of the
materials near to our bolometers, bulk contaminations of materials
both near and far from our bolometers, and cosmogenic muons. More
information about the \Q \MC simulation can be found in
\cite{CUOREBudget,Q0BackgroundRecon}.

\begin{figure*}
  \centering
  \includegraphics[width=.9\textwidth]{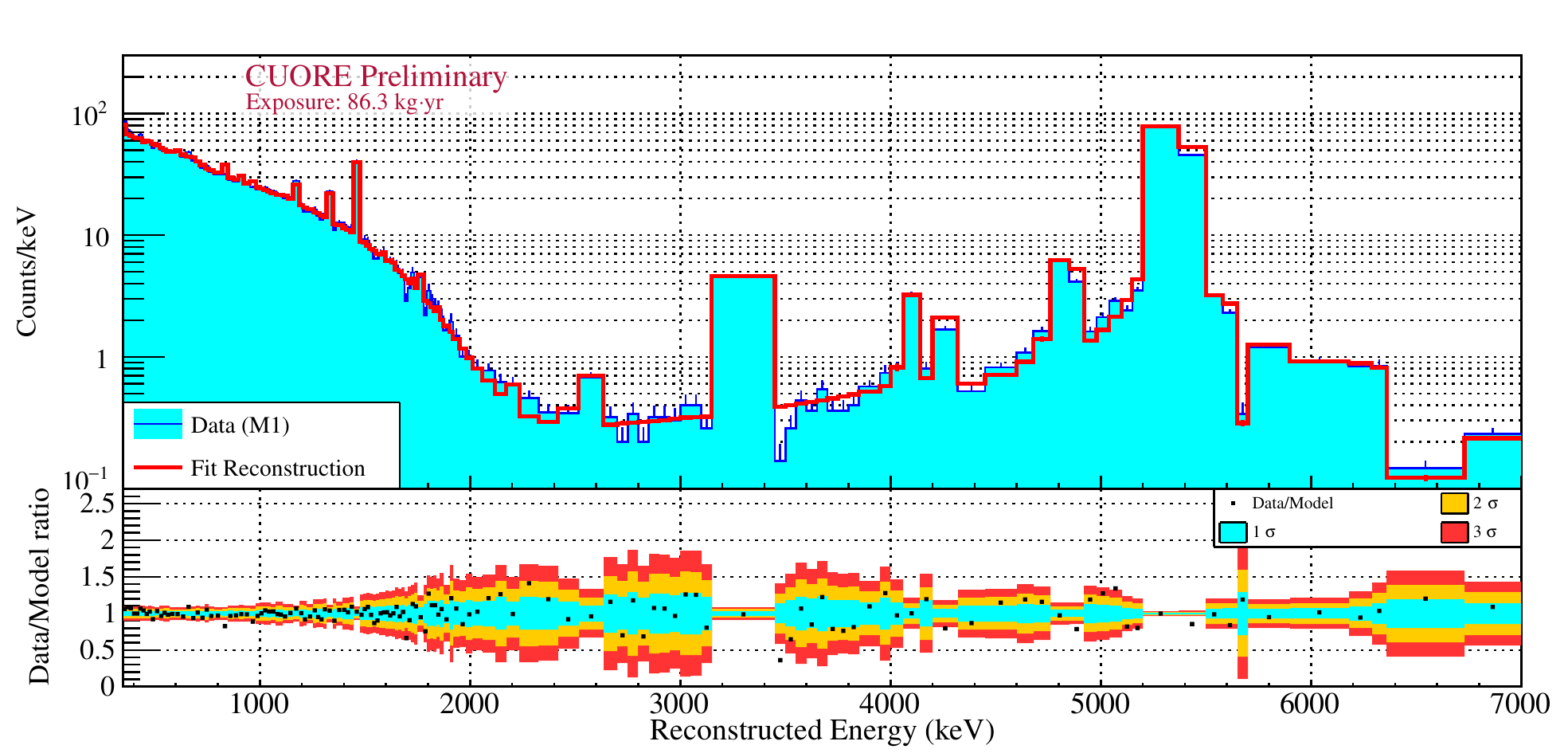}
  \caption{Top: The measured M1L0 spectrum (\emph{blue}) and its
    reconstruction (\emph{red}). The spectra are binned with an
    adaptive binning to contain peaks into a single bin (to avoid
    dependence on the peak shape), while also achieving good
    resolution of the continuum shape. Bottom: The ratio of the data
    to the reconstructed model with 1$\sigma$, 2$\sigma$ and 3$\sigma$
    error bars. It is clear from the data that we are able to
    faithfully reconstruct the continuum, with moderate disagreement
    in the heights of a few peaks.}
  \label{fig:M1L0}
\end{figure*}

\begin{figure*}
  \centering
  \includegraphics[width=.9\textwidth]{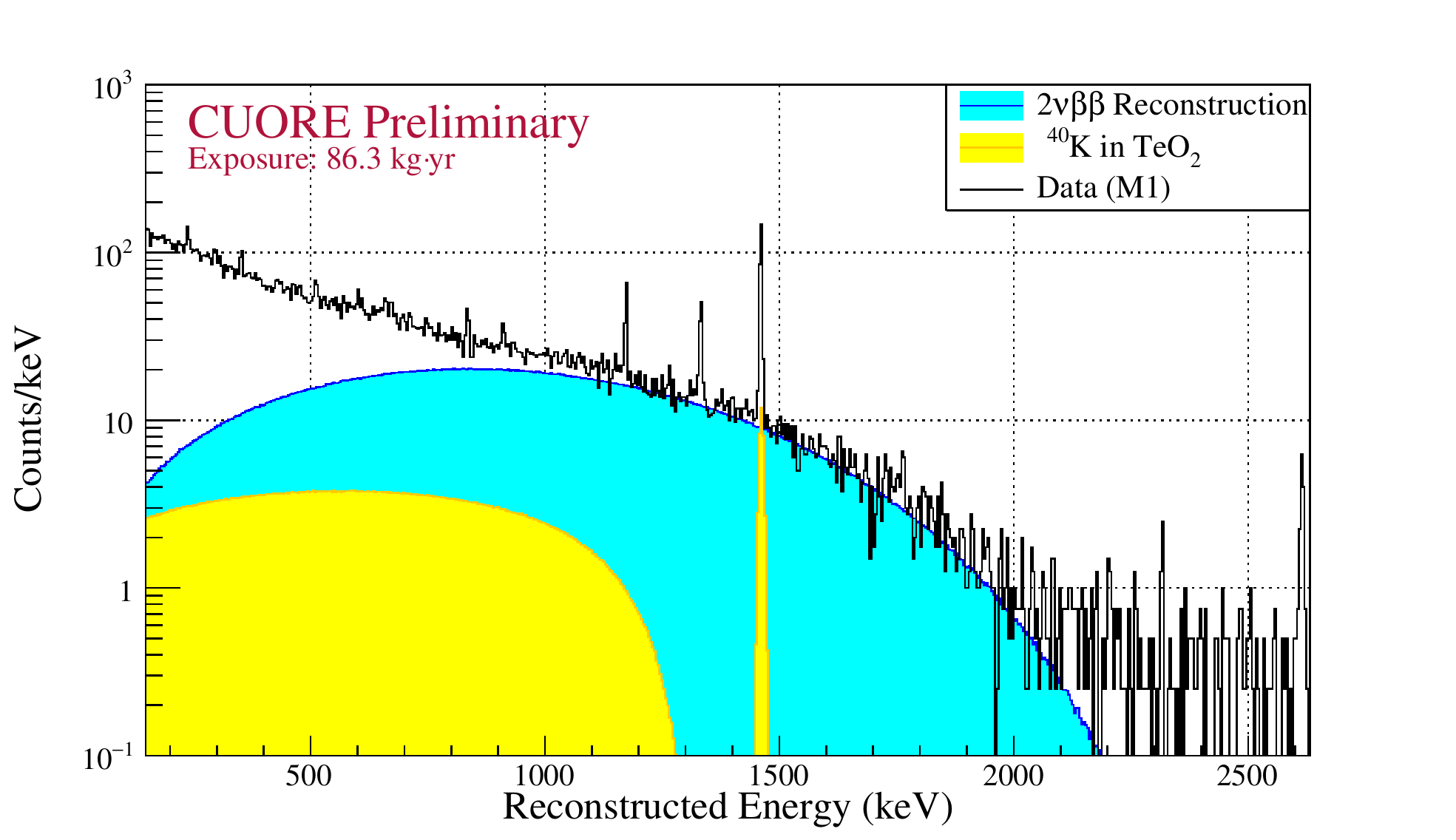}
  \caption{The observed M1L0 spectrum (\emph{black}) with a
    reconstruction of the 2\nbb component (\emph{blue}) of the
    background as well as the $^{40}$K background (\emph{yellow}). The
    2\nbb component dominates the observed M1L0 spectrum between
    $\sim$1 to $\sim$2\,MeV. Note: The observed spectra has been
    converted back to the 1\,keV binning and the 2\nbb and $^{40}$K to
    smooth continua for illustrative purposes, the fit is performed
    with the binning in Fig.~\ref{fig:M1L0}.}
  \label{fig:2nbb_40K}
\end{figure*}

We split the data into three types of spectra: a multiplicity 1 (M1)
spectrum comprised of events where energy was deposited into a single
bolometer, a multiplicity 2 (M2) spectrum comprised of the single
bolometer energies of events where energy was shared between two
bolometers, and an M2 sum spectrum ($\Sigma$2) comprised of the summed
energy of the M2 events. Double beta decay events deposit all of their
energy into a single bolometer about 90\% of the time and so are
primarily in the M1 spectrum, whereas many of our backgrounds deposit
energy across two or more bolometers (e.g.\ $\gamma$-rays that scatter
from one crystal into another or $\alpha$-decays that occur on a
surface between two neighboring crystals.) So the M2 and $\Sigma$2
spectra are particularly useful for understanding our backgrounds (see
Fig.~\ref{fig:M1_vs_M2Sum}). The M1 spectrum is further split into two
independent spectra: the layer 0 spectrum (M1L0) which is comprised of
the 252 ``inner core'' bolometers, which are expected to be more
shielded from external backgrounds; and the layer 1 spectrum (M1L1)
comprised of the 736 bolometers on the surface of the detector, which
are expected to be more susceptible to backgrounds originating outside
the detector.

We then reconstruct the observed \Q background by fitting the \MC
simulated spectra to the observed data simultaneously across these
four spectra using a \MCMC implemented in the JAGS software package
\cite{JAGS1,JAGS2,JAGS3}. The fit has a total of 60 free parameters
corresponding to the contamination levels of each background
component. Both the observed and \MC spectra are binned with variable
bin sizes to reduce the effect of the complicated line shapes. The
fitting procedure closely follows the procedure laid out in
\cite{Q0BackgroundRecon} and a paper describing the precise procedure
for \Q is in preparation.

\begin{table}
  \centering
  \caption{Table of fit $\chi^2$ values for the four spectra, along with the associated number of degrees of freedom.}
  \label{tab:chi2}
  \begin{tabular}{lccc}
    \hline
    \hline
    & $\chi^2$ & d.o.f. & $\chi^2_{\rm red}$\\
    \hline
    M1L0 & 388.5 & 139 & 2.80 \\
    M1L1 & 387.5 & 158 & 2.20\\
    M2 & 408.5 & 130 & 3.15 \\
    $\Sigma2$ & 273.9 & 102 & 2.69 \\
    \hline
    Total & 1418.7 & 529 & 2.68\\
    \hline
    \hline
  \end{tabular}
\end{table}

Overall the model is able to reproduce nearly all of the major
features of the observed spectra (see Fig.~\ref{fig:M1L0}). The
$\chi^2$ values are reported in Table~\ref{tab:chi2}, and show
reasonable agreement between the fit and the data. Much of the
disagreement between the data and reconstruction comes from a poor
reconstruction of several $\alpha$ and $\gamma$ peaks which -- due to
the high statistics of the peaks -- have a large effect on the
$\chi^2$. But practically, these indicate that we have not yet
properly modeled all the contaminations leading to these
backgrounds. In the future, effects such as these will improve with
increased statistics as we are able to disentangle the origins of the
contaminations.

\begin{figure}
  \centering
  \includegraphics[width=.49\textwidth]{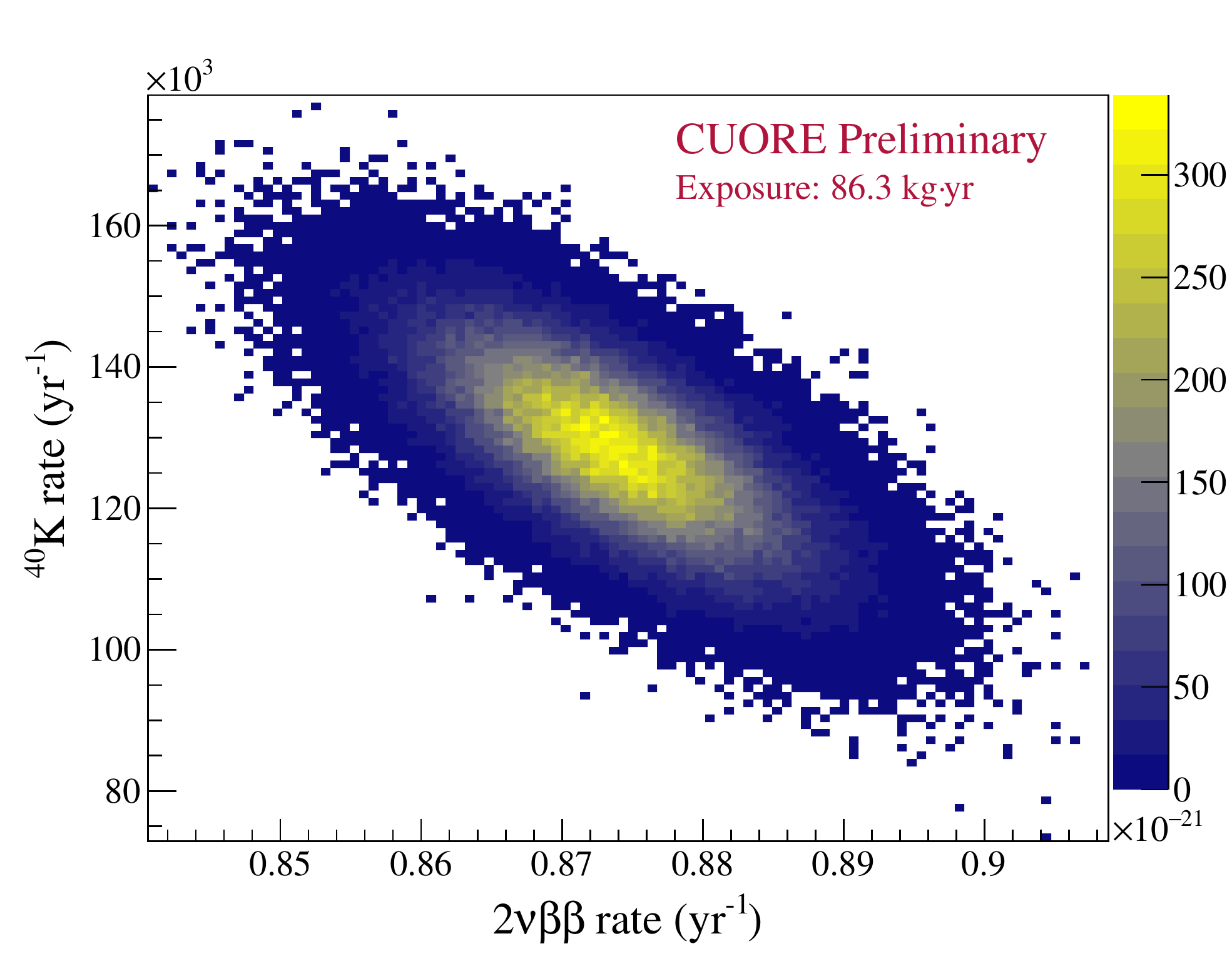}
  \caption{The correlation between the reconstructed 2\nbb decay rate,
    $\Gamma^{2\nu}$ and the observed $^{40}$K event rate. The
    anti-correlation with $^{40}$K in the \TeO bulk is the largest
    statistical correlation observed.}
  \label{fig:corrmatrix}
\end{figure}

The remaining systematic effects create a very large uncertainty in
the background rate in the \NDBD \ROI around \Qbb. As such, we leave a
measurement of the background index and its effect on the sensitivity
of \Q to \NDBD for future work.

Despite these systematic uncertainties, we are able to extract a
robust estimate of the 2\nbb half-life. Due to the very low
$\gamma$-backgrounds and increased \isoTe mass of \Q, the 2\nbb decay
is the dominant component of the observed M1 spectrum from about 1 -
2\,MeV (see Fig.~\ref{fig:2nbb_40K}). To prevent bias while tuning our
data quality cuts and fitting procedure, we blinded our \MC
normalization constant to keep the extracted half-life in terms of an
unphysical ratio that could not be compared to previous results. When
the fit procedure was finalized, we unblinded the correct
normalization and extracted a measurement of the 2\nbb half-life of
\isoTe of
\mbox{$T_{1/2}^{2\nu}=\left[7.9\pm0.1\,\mathrm{(stat.)}\pm0.2\mathrm{(syst.)}\right]\times10^{20}$\,yr}. This
result is consistent with previous measurements of this half life (see
Table~\ref{tab:halflives}). The present result represents the most
precise measurement of the 2\nbb half-life of \isoTe to date and one
of the most precise measurements of a 2\nbb decay half-life to date.

\begin{table}
  \centering
  \caption{Recent measurements of the $T_{1/2}^{2\nu}$ of \isoTe.}
  \label{tab:halflives}
  \begin{tabular}{lccc}
  \hline
  \hline
  & $T_{1/2}^{2\nu}$ ($10^{20}$\,yr) & Frac.Uncert. & Ref. \\
  \hline
  MiBeta & $6.1\pm1.4 \,{}^{+2.9}_{-3.5}$& 57.3\% & 2003 \cite{MiDBD2nu} \\
  NEMO-3 & $7.0 \pm 0.9 \pm 1.1$ & 20.3\% & 2011 \cite{NEMO3Te2nu} \\
  CUORE-0 & $8.2\pm0.2\pm0.6$ & 7.7\% & 2016 \cite{Q0BackgroundRecon}\\
  CUORE  & $7.9\pm0.1\pm0.2$ & 2.8\% & (this result) \\
  \hline
  \hline
  \end{tabular}  
\end{table}

The only component of the reconstructed model that is strongly
correlated with the 2\nbb decay rate is the contamination level of
$^{40}$K in the \TeO crystal bulk (see
Fig.~\ref{fig:corrmatrix}). $^{40}$K has a $\beta^-$-decay with a
1310.9\,keV endpoint (see Fig.~\ref{fig:2nbb_40K}), which becomes
correlated with the broad 2\nbb spectrum. A similar effect was seen in
\q which had an identical crystal growth process and the same $^{40}$K
contamination levels. But the larger statistics of \Q and the lower
background -- particularly in the 1 - 2\,MeV region -- makes this only
a small ($\sim$1\%) statistical uncertainty on the final measured
rate.

One of the major improvements over the first \Q \NDBD result, is an
improvement to the estimation of the signal acceptance
efficiency. Previously, we had been using several $\gamma$ lines to
estimate the efficiency of good signal events passing the pulse shape
cuts to contribute to the final observed spectrum. This approach was
limited by statistics and had to be interpolated between $\gamma$
lines and thus contributed a $\sim$2.4\% uncertainty to the final
result. In the present analysis, we are using the M2 spectrum. Due to
the very low event rate of \Q, accidental M2 events are rare
($\lesssim 1\%$) and so the M2 spectrum constitutes a clean sample of
events, distributed in energy, to test for the signal efficiency. This
significantly increases the statistics relative to the previous
approach and as a result decreases the uncertainty on the present
signal efficiency to $\lesssim$0.2\%.

The leading systematic uncertainty on the measurement of
$T^{2\nu}_{1/2}$ is now the geometric splitting of the data into
subgroups. Our fit reconstruction splits the data into an inner layer
of 252 bolometers and an outer layer of 736 bolometers, however, other
splittings of the data are also possible and yield different
results. Other splittings include looking at only even or odd
channels, even or odd towers, even or odd floors, the top half vs
bottom half of the detector, etc. These different splittings of the
data by geometry allow us to probe the uncertainty caused by our
present ignorance of the exact origin of localized contaminations. The
precise details of this procedure will be laid out in future
publications.

\section{Cryostat Maintenance}

After the two datasets collected in summer 2017, we paused data taking
for a period of optimization and maintenance on the \Q cryostat. This
began with a scan of the detector performance of a function of
temperature. We measured the noise levels and energy resolution of
heater pulses at operating temperatures of 11, 13, 15, 17 and
19\,mK. Based on this data, we selected a new operating temperature of
11\,mK, as opposed to the 15\,mK at which we had previously been
operating.

Starting in January 2018, we warmed the detector to 100\,K to upgrade
a set of vacuum gate valves. This operation was performed without
fully warming to room temperature, but warm enough to allow us to
overpressure the vacuum region with He gas. In total, this operation
took about 2 months to complete, which is very quick given the size
and cold mass of the \Q cryostat.

We returned to base temperature in March 2018 and performed a scan of
the pulse tubes to find the optimal phase configuration of the 4 \Q
pulse tubes that minimizes the noise on the detector (see
\cite{PTPhaseScan}). In April 2018, we returned to our operating
configuration and performed a detector calibration. We were able to
return to our previous data taking configuration, recovering a
7.6\,keV \FWHM energy resolution and continue working on improve the
detector operation. We resumed stable physics data taking in May 2018.

\section{Conclusion}

\Q began collecting data in summer of 2017 and, with the first two
datasets and an exposure of 24.0\,\kgyr of \isoTe exposure, has
already set the strongest limit on the \NDBD decay half-life of \isoTe
to date at \mbox{$T_{1/2}^{0\nu}>1.5\times10^{25}$\,yr} at \mbox{90\%
  C.L.} Recently, we have been working to reconstruct the background
and were able to make a measurement of the 2\nbb decay half-life of
\isoTe of
\mbox{$T_{1/2}^{2\nu}=\left[7.9\pm0.1\,\mathrm{(stat.)}\pm0.2\,\mathrm{(syst.)}\right]\times10^{20}$\,yr.}
This is the most precise measurement of the 2\nbb decay half-life of
\isoTe and one of the most precise measurements of any 2\nbb decay to
date.

In the past few months, \Q has undergone a period of upgrades and
optimization and has entered a period of stable data taking. We
continue to work to improve the energy resolution to our goal of
5\,keV \FWHM and reach the background goal of
\mbox{$b=0.01$\,\ckky}. The ultimate sensitivity goal for \Q is
\mbox{$9.0\times10^{25}$\,yr} after 5 years of live-time.

\bibliography{biblio}

\end{document}